\documentstyle[11pt,newpasp,twoside,psfig]{article}
\index{summary}
\index{instructions}
\index{template}

\def\edcomment#1{\iffalse\marginpar{\raggedright\sl#1\/}\else\relax\fi}
\reversemarginpar

\begin{document}
\title{X-ray Spectrum and Pulsations of the Vela Pulsar}
\author{Divas Sanwal, G.~G.~Pavlov, O.~Y.~Kargaltsev, G.~P.~Garmire}
\affil{Penn. State University, 525, Davey Lab, University Park, PA 16802}
\author{V.~E.~Zavlin, V.~Burwitz}
\affil{Max-Planck-Insitut f\"ur Extraterrestrische Physik, Garching, Germany}
\author{R.~N.~Manchester}
\affil{ATNF, CSIRO, P.O. Box 76, Epping, NSW 1710, Australia}
\author{R.~Dodson}
\affil{School of Maths \& Physics, University of Tasmania, Australia}

\begin{abstract}

We report the results of the spectral and timing analysis of
observations of the Vela pulsar with the {\sl Chandra}
X-ray Observatory.
The spectrum shows no statistically significant spectral
lines in the observed 0.25--8.0 keV band.  It consists of two
distinct continuum components. The softer component can be
modeled as either a magnetic hydrogen atmosphere spectrum
with $kT_{{\rm eff},\infty} = 59\pm 3$ eV,
$R_\infty =15.5\pm 1.5$ km, or a standard blackbody with
$kT_\infty =129\pm 4$ eV, $R_\infty = 2.1\pm 0.2$ km (the
radii are for a distance of 250 pc).  The harder component,
modeled as a power-law spectrum, gives photon indices
depending on the model adopted for the soft component:
$\gamma=1.5\pm 0.3$ for the magnetic atmosphere soft component,
and $\gamma=2.7\pm 0.4$ for the blackbody soft component.
Timing analysis shows three peaks in the pulse profile,
separated by about 0.3 in phase.  Energy-resolved timing
provides evidence for pulse profile variation with energy.
The higher energy ($E > 1.8$ keV) profile shows significantly
higher pulsed fraction.

\end{abstract}

\vspace{-0.6cm}
\section{Introduction}

The Vela pulsar (B0833-45), with a period of 89 millisecond and
a characteristic age of 11,000 years,
it is the prototype of the young  ``Vela-like'' pulsars.
The excellent angular resolution of  {\em Chandra} allows us to
separate the pulsar from its synchrotron nebula and study its
X-ray properties.

\vspace{-0.6cm}
\section{Spectrum}

We extracted a total of 16000 counts in the range 0.25--2.0 keV
from two HRC-S/LETG observations of 25~ks each.
No deviations of counts in
the individual bins from the mean value of neighboring 10--20
bins were  found  with statistical significance higher than
2.7$\sigma$. This may indicate that there are no elements
heavier than H present on the NS surface.  The spectral
fits with thermal models show excess counts at the
high energy end.

To get a handle on the harder component, we used the publicly
available ACIS/HETG data obtained in the Continuous
Clocking (CC) mode.
We used only the zero order image, which is equivalent to about
3~ks observation without the grating.  We applied manual corrections
for the dither and the SIM motion for both timing and the event
position in sky plane (see Zavlin~et~al. 2000 for details).  The
background was estimated by
interpolating between the neighboring regions in the 1D image of
the pulsar and its nebula. Since the
ACIS response is poorly known at E $<$ 0.5 keV, and the
background dominates above 8.0 keV in the CC mode, we use
the events in the energy range 0.5--8.0 keV.

\begin{figure}[t]
\begin{center}
\vspace{-0.8cm}
\hbox{
\psfig{figure=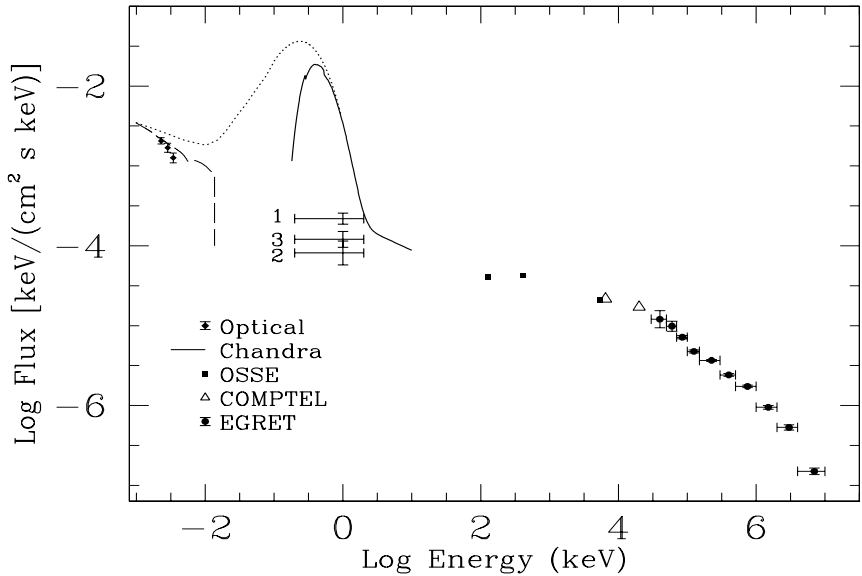,height=8.0cm,width=12.0cm,angle=270}
}
\vspace{-1.2cm}
\end{center}
\caption
{ The multiwavelength spectrum of the Vela pulsar. The data shown include
optical (Nasuti~et~al. 1997), OSSE (Strickman~et~al. 1996), COMPTEL
(Sch\"onfelder~et~al. 2000) and EGRET (Kanbach~et~al. 1994). The soft X-ray
flux corresponds to the NS atmosphere model for the thermal component. The
dotted line is the unabsorbed flux while the solid line is the observed
flux from the Vela pulsar.  The pulsed fluxes corresponding to the
three HRC peaks (see Fig. 2) are marked with horizontal bars. }
\label{fig1}
\vspace{-0.5cm}
\end{figure}

\begin{table}[ht]
\vspace{-0.8cm}
\begin{center}
\caption{Parameters of the two-component fits to the Vela pulsar spectrum. The
radii and luminosities are for a distance of 250 pc.}
\begin{tabular}{lcccccc}
\tableline
Thermal & $n_H$ & T$_{\rm eff}$ & R$_{\rm eff}$ & L$^{\rm th}_{\rm bol}$ & $\gamma$ & L$^{\rm nonth}_{\rm 0.5-8.0 keV}$ \\
Model   & ($10^{20}$ cm$^{-2}$) & (MK) & (km) & (erg s$^{-1}$) & & (erg s$^{-1}$) \\
\tableline
Blackbody & 2 & 1.5 & 2 & 1.5$\times 10^{32}$ & 2.7 & $2.0 \times 10^{31}$ \\
H Atmosphere &3 & 0.7 & 12 & 3.8$\times 10^{32}$ & 1.5 & 1.5$\times 10^{31}$ \\
\tableline\tableline
\end{tabular}
\vspace{-0.5cm}
\end{center}
\end{table}

The combined spectral fit to the HRC-S/LETG and ACIS/HETG-CC
data clearly shows two components -- a soft thermal component,
which fits equally well with a blackbody or a magnetic
neutron star atmosphere model,
and a harder component, which we interpret as a power-law (PL).
The parameters of the fit depend on the model chosen for the soft
thermal component.  Table~1 shows the parameters of fits to the
combined data (see Pavlov~et~al. 2001 for details) . Figure~1
shows the broadband spectrum of the Vela pulsar.  The soft X-ray
spectral parameters are for the atmosphere model.  The
extrapolation of the PL component matches both the optical and
the hard X-ray/$\gamma$-ray  flux.
This is the first conclusive observation of both the thermal and
the non-thermal radiation from the Vela pulsar.

\vspace{-0.6cm}
\section{Timing}

The Vela pulsar was observed twice with the HRC-I and
twice with the HRC-S/LETG. Both of the HRC-I observations
and the first LETG observation
suffered from the HRC timing problem. We corrected the
times by shifting to the time of the next event in the
level~1 event list taking only the events with time
correction less than 4 ms. The second LETG observation
was in the timing mode, so we could recover all the
event times.

\begin{figure}[t]
\begin{center}
\vspace{-0.8cm}
\hbox{
\psfig{figure=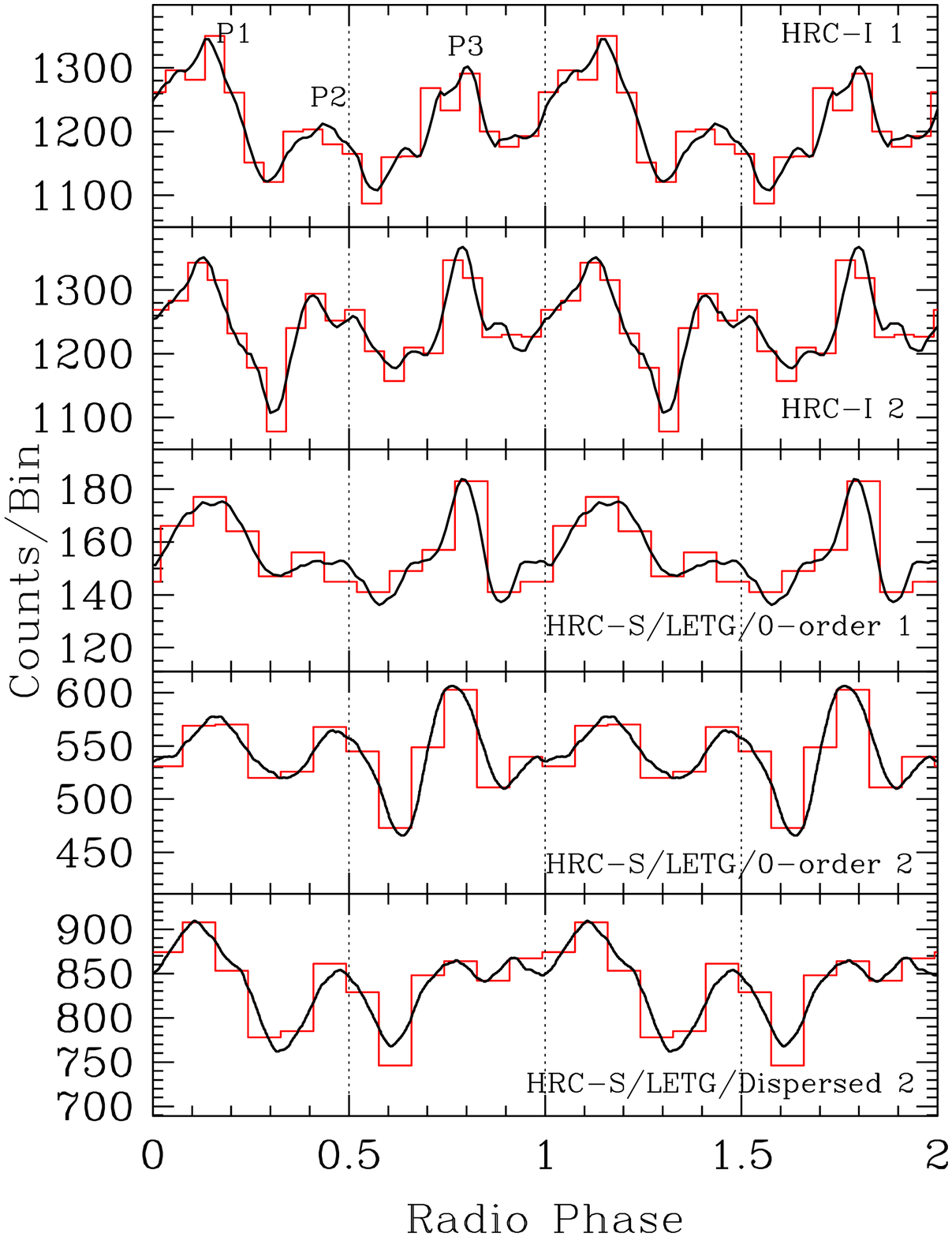,height=10.0cm,width=7.0cm,angle=0}
\hspace{-0.7cm}
\psfig{figure=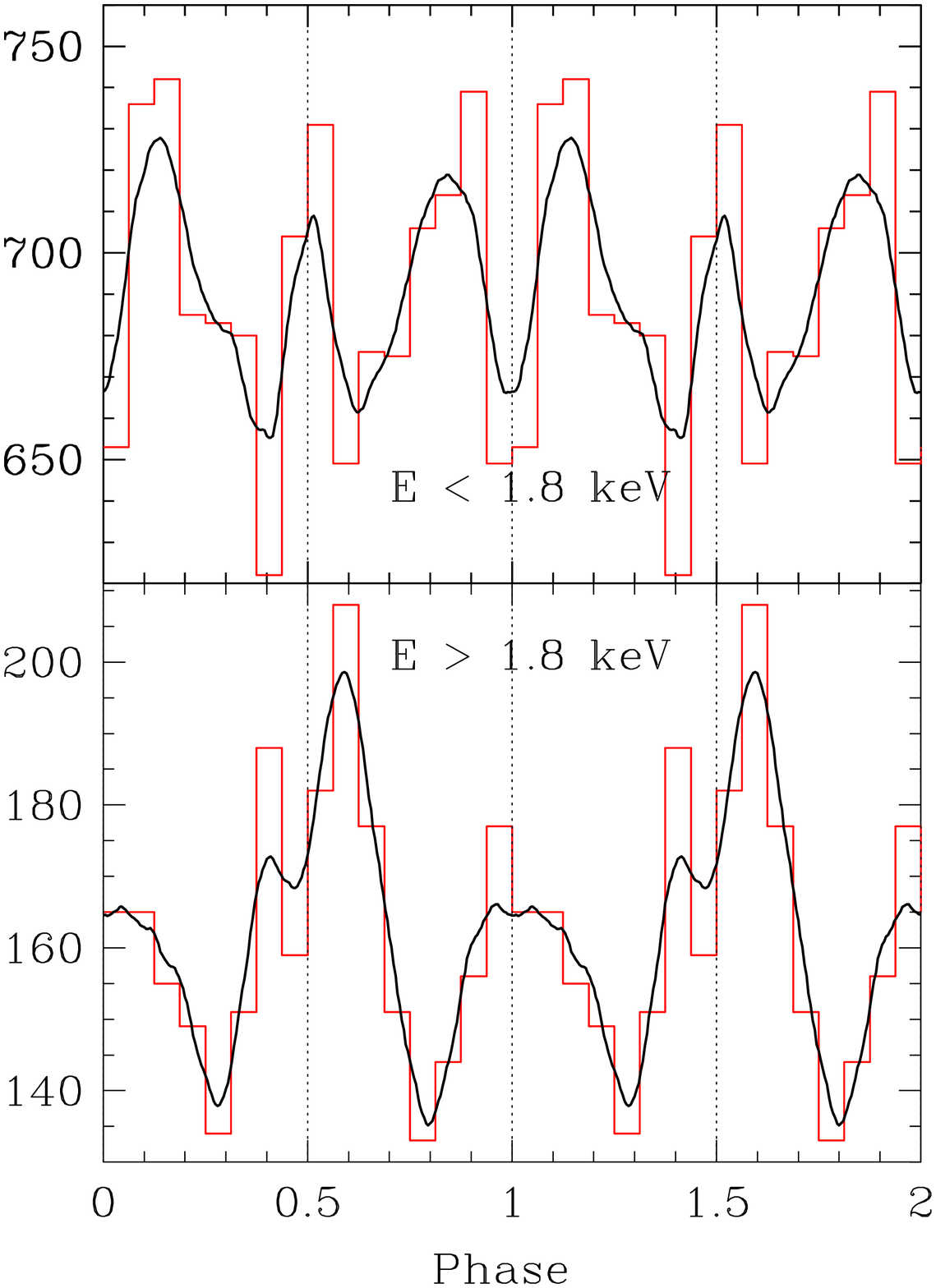,height=10.0cm,width=6.8cm,angle=0}
}
\vspace{-1.0cm}
\end{center}
\caption{
{\bf {\it Left}}: The pulse profiles from the HRC
observations. The three peaks (P1, P2 and P3) are marked in the top panel.
{\bf{\it Right }}: Energy-resolved pulse profiles from the ACIS/HETG-CC
observation.  The top panel shows the pulse profile
for $E < 1.8$ keV corresponding to the thermal component dominated regime.
The bottom panel shows the profile for $E > 1.8$ keV, where 
non-thermal component dominates.}
\vspace{-0.2cm}
\label{fig2}
\end{figure}

We used the radio ephemeris of the Vela pulsar based on the
observations at the Hobart and Parkes radio observatories.
The phase of each event was determined and
folded to get the pulse profiles.  The left panel of Figure~2
shows the observed pulse profiles of the HRC observations.  The
pulse profiles show three peaks separated by about 1/3
of phase, with  a total pulsed fraction of $9\%$ (see also
Helfand~et~al. 2001).  The pulsed fluxes in the three peaks
are shown in Figure~1.

For the ACIS/HETG-CC observation, the recorded times are the event
readout times.  We corrected the times for the dither and SIM
motion, and the delay from the event to its readout. The correction
for absolute time depends on the actual position of the source on
the CCD, which is poorly known at present.  Therefore, we are unable
to get the absolute times for the CC observation.
The relative times, on the other hand, are accurate.  The right
panel of Figure~2 shows the ACIS pulse profiles
in two energy bands ($E < 1.8$ keV and $E > 1.8$ keV).
The difference in the profile shapes and the pulsed fraction
is striking.  We estimate
the intrinsic pulsed fraction of the pulsar to
be $8\pm2\%$ and $62\pm20\%$ in the $E < 1.8$ keV and
$E > 1.8$ keV bands, respectively.


\vspace{-0.3cm}
\section{Summary}

\begin{itemize}

\item The HRC-S/LETG spectrum shows no conclusive evidence for lines
in the 0.25--2.0 keV range.
The spectrum shows two components (thermal + non-thermal), similar to
middle-aged pulsars. The thermal component is consistent with the
emission from a magnetic Hydrogen atmosphere.  Extrapolation of
the PL component matches the optical and the hard X-ray points.

\item The thermal luminosity of the Vela pulsar is lower than that
predicted by the ``standard'' neutron star cooling models (Tsuruta 1998).

\item HRC data show three peaks with total pulsed fraction of about $9\%$.
Energy-resolved ACIS pulse profiles show energy-dependent shape and
pulsed fraction.  The pulse profile at $E < 1.8$ keV is similar to
that observed with HRC.  At $E > 1.8$ keV, the pulse profile shows
only two peaks with estimated intrinsic pulsed fraction of $62\pm20\%$.

\end{itemize}

\acknowledgments

This work was supproted by NASA grants NAG5-7017, NAG5-10865, NAS8-3852, and
SAO grant GO1-2071X.

\end{document}